\documentclass{article}
\usepackage{rfia2000}
\usepackage{color}
\usepackage[latin1]{inputenc}
\usepackage{graphicx}
\usepackage{url}
\usepackage{times}
\usepackage{hyperref}

\begin{document}
\title{The WebContent XML Store \thanks{This research was
supported by the French National Research Agency (ANR) through the RNTL program, and the System@tic Paris-R\'egion cluster.}}

\author{Benjamin Nguyen\\PRISM -- UVSQ \\Project SMIS -- INRIA \and Spyros
Zoupanos\\Project LEO -- INRIA Saclay - Île-de-France\\ University of Paris-Dauphine}

% negative vspace helps get keywords in left column

%\additionalauthors{\ep{contributors who could not fit on the 1st page will go there}}
\date{}

\maketitle
\section*{Abstract}

In this article, we describe the XML storage system used in the WebContent
project. We begin by advocating the use of an XML database in order to store
WebContent documents, and we present two different ways of storing and querying
these documents : the use of a centralized XML database and the use of a P2P
XML database.

%fin de l'abstract

\section{Context}
\label{sec:overview}
\paragraph{Overview:}
The WebContent platform~\footnote{\url{http://www.webcontent-project.org/}}
proposes a specific UML schema to be used by all its services. Through a canonical transformation, this schema can
be converted into an XML Schema. This is extremely usefull
since the Web Services paradigm uses XML documents to communicate
with each other. It seems therefore straightforward to manage \emph{all} the
documents inside the WebContent platform in XML format, which will present
advantages when storing them and querying them. In this article, we describe
two ways of managing the storage and querying of such documents, by using a
centralized and distributed (P2P) XML database. These
\emph{storage-service} modules conform to the WebContent interface for
storage. The main reason for chosing to use XML
databases over a simple file storage format is twofold : performance and
expressivity of queries, since as we will see, it is possible to express any
sort of XQuery on a WebContent document.
\paragraph{WebContent Storage Services Interface:}
The platform defines an interface for a \emph{storage} service and
consequently a \emph{query} service, to access the data that is
stored. These interfaces are generic. To illustrate their flexibility,
we have provided two
implementations~\cite{WebContentDemoWorkshop, WebContentDemoVLDB}. The first
one provides storage and querying on top of existing single-site (centralized)
XML database servers using an existing XML Query engine :
MonetDB~\footnote{\url{http://monetdb.cwi.nl/XQuery}}. Second, we have
implemented a resource store distributed over the network peers, and similarly a query service implemented
jointly by all the peers in the network. We stress that moving from one
implementation of the storage service to another is totally transparent to the
user, and similarly for the query service.

\section{Storage services} 
\paragraph{Centralized Store:} For storage on a single machine, we
can use either  MonetDB or MS SQL
Server. In both cases, the WebContent
documents are stored in their native XML format, and can be queried
via XQuery. An issue with such queries is that they may return
results of any (XML) type. Therefore, we have developed a specific WebContent
query interface that only allows queries returning WebContent
resources, which may be placed in the warehouse.

\paragraph{P2P storage service:} The P2P
storage service is implemented jointly by several peers, so that 
the exact location
of a piece of data is transparent to the user.  The P2P storage
service also supports indexing facilities.  A {\em DHT service} is
implemented on top of a distributed hash table (or DHT, for
short~\cite{DHTAPI}). The DHT, a distributed software running on all peers, provides
the connectivity of the network. It assigns unique identifiers to
peers and allows them to easily join and leave the
network\footnote{Remember that in a hybrid architecture, all
participants need not be part of the P2P network.}.  Indexing is
supported using a distributed data structure based on the simple
abstraction of (key,value) pairs (with two services, namely {\em
put(k,v)} and {\em get(k)}).  Without delving into the details, the
DHT stores all values associated to a given key $k$, on a particular
peer in charge of that key.

Different DHTs may have different algorithmic properties, interesting
from different performance viewpoints. For instance, a DHT may
guarantee that two keys $k_1$ and $k_2$, ``close'' by some distance
measure, are managed by peers that are ``close'' in some sense. To
take advantage of the good properties of distinct DHTs, several DHTs
may coexist in a WebContent deployment architecture. Thus, a peer $p$
belonging to the DHTs $dht_1,dht_2,\ldots$ is an endpoint for the
services $join_1,leave_1,put_1,get_1$, but also for
$join_2,leave_2,put_2,get_2$ etc.  We have successfully integrated so
far two DHTs~\cite{WebContentDemoVLDB}: FreePastry~\cite{Pastry} from
MIT, including our own extensions for robust scalable XML
indexing~\cite{ICDE2008}; and PathFinder~\cite{PathFinder}, specially
tuned to support interval search queries (which FreePastry does not
support). The Active XML~\footnote{\url{http://www.activexml.net}} engine is
responsible to interact with the available DHTs since their presence and query
processing performed by each of them should be transparent to the user.

\section{Query services}

\paragraph{XQuery:}
WebContent resource exploitation relies on advanced query processing
capabilities. To this end, we use {\em XML query services}. In its
centralized (one-site) implementation, an XML query service takes as
input an XQuery~\footnote{\url{http://www.w3.org/TR/xquery/}}, and returns its
results as evaluated by the underlying XML DB. Observe that in
this context, it is only meaningful to solicit the query service on
the machine that stores the queried document(s). XQuery is an extremely powerful
language, and it is possible to write many complex queries in particular to
restructure or perform joins on the documents. While the WebContent interface
allows such queries to be written, the main functionality of the store is to
provide access to any \emph{resource} that is stored in the database. Recall
that a resource can be anything from a document to one of its atomic resources
such as a paragraph. Such queries are much simpler than XQueries, and are
implemented in the centralized service using an index on all the resource
elements. It is therefore possible to find in time O(1) any resource stored
in the database (serialization cost is of course function of the size of the
resource).

\begin{figure}
\begin{center}
\includegraphics[width=\columnwidth]{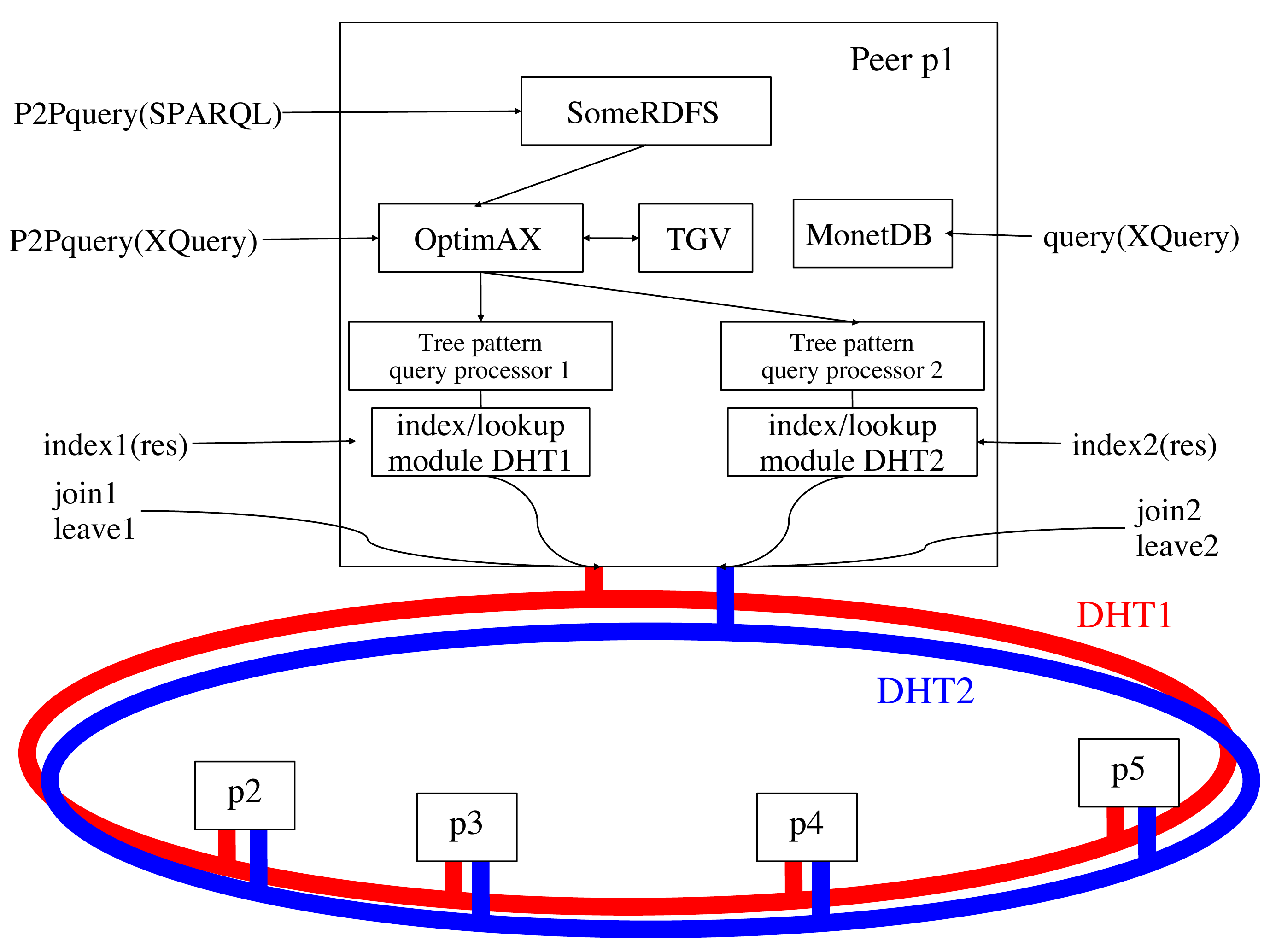}
\caption{Outline of P2P services.\label{fig:p2p}}
\end{center}
\end{figure}

\paragraph{P2P Query engine:}The implementation of our P2P XML query service is
more evolved. This service is provided by any WebContent peers, and is implemented by
several collaborating peers. The queries it supports may be asked 
against the set of all documents available in the warehouse,
regardless of their location. The processing of such a query can be
traced on the Figure~\ref{fig:p2p}. This figure shows a P2P WebContent
network based on two superposed DHTs, such as KadoP and PathFinder
which we integrated. Accordingly, the detailed
structure of peer $p_1$ features a tree pattern query processor for 
each of the DHTs. Classical database optimization
techniques can be incorporated into each of these tree pattern query
processors, e.g., a query cache has been built in
the KadoP tree pattern query processor etc.

The query is handled to a
{\em P2P optimizer service} we developed, namely
OptimAX~\cite{OptimAX}, which performs two tasks. ($i$)~Based on the
knowledge it has of the available DHT indices, and with the help of an
embedded XQuery algebraic compiler~\cite{TGV}, OptimAX extracts from
the query: the maximal subqueries that can be processed by the
each available tree pattern query processors, and a
recomposition query which assembles the results of index lookups into
the desired query result form. ($ii$)~The calls to the KadoP and/or
PathFinder indexes are placed in the network of peers in such a way as
to reduce the total amount of data transfers incurred by query processing.
OptimAX is implemented as a rule-based rewriting engine and execution
plans are encoded as ActiveXML documents. Once OptimAX has produced an
execution plan, it is given to the AXML engine for execution. This is carried out by relying on the tree pattern query capabilities of KadoP~\cite{ICDE2008} and
PathFinder~\cite{PathFinder}, and on the XML query service local to
the query peer for the recomposition query. 

Finaly, each peer is capable of precessing semantic queries over RDF
data, expressed in a conjunctive subset of the
SPARQL~\footnote{\url{http://www.w3.org/TR/rdf-sparql-query/}} language.

\bibliographystyle{abbrv}
\bibliography{rfia} 
\end{document}